\documentclass[reprint,
bibnotes,
amsmath,amssymb,
aps,
]{revtex4-2}
\usepackage{caption}
\usepackage{subcaption}
\usepackage{graphicx}
\usepackage{dcolumn}
\usepackage{bm}

\begin{document}
\title{Current induced by a tilted magnetic field in phosphorene under terahertz laser radiation}

\author{Narjes Kheirabadi }
\affiliation{Department of Physics, Sharif University of Technology, Tehran 14588-89694, Iran}%


\begin{abstract}
In this study, we investigate the cyclotron resonance effect in the first-order AC current, magnetic ratchet effect, and second harmonic generation in phosphorene in the presence of a steady tilted magnetic field and under THz laser radiation. We establish that various cyclotron resonances exist in the deduced currents based on the angular frequency of the incoming light. These resonances are dependent on the $\omega_{c_{+}}$ value, a function of the carrier charge, the perpendicular magnetic field, and the effective masses along the armchair or zigzag edges. We discuss the direction and the magnitude of the deduced currents for various radiation polarizations. We compare the results with a zero perpendicular magnetic field. Cyclotron resonance for the first order AC current occurs at $\omega= \pm \omega_{c_{+}}$. The deduced current declines if the perpendicular magnetic field is zero. Meanwhile, for the ratchet current, cyclotron resonances occur at $\omega= \pm \omega_{c_{+}}$, $\omega= \pm 2\omega_{c_{+}}$, and radiation helicity affects the deduced current for circularly polarized light. Cyclotron resonance for the second harmonic generation current occurs at $\omega= \pm \omega_{c_{+}}$, $\omega= \pm 2 \omega_{c_{+}}$ and $\omega= \pm \omega_{c_{+}}/2 $ and the current is stronger compared to the case with no perpendicular magnetic field. As the magnetic field rotates in the plane of anisotropic phosphorene, separate directions are predicted for the second harmonic generation--related current. It is noteworthy that the magnitude of the ratchet and second harmonic generation current are within the same range and comparable to the magnetic ratchet current in monolayer graphene, $\mu A / cm$. 
\end{abstract}

\maketitle
\section{Introduction}
Since its discovery in 2004, graphene research has observed exponential growth and has founded a new era in nanomaterials, and their applications \citep{novoselov2004electric, geim2010rise,novoselov2012roadmap}. The revolutionary discovery has triggered a pursuit for two--dimensional (2D) materials beyond graphene \citep{butler2013progress}. Subsequently, a series of monolayer 2D materials have been produced from graphene to transition metal dichalcogenides (TMDCs), and phosphorene \citep{huo20152d, shim2017electronic, akhtar2017recent, peruzzini2019perspective}. These materials have diverse electronic and optical properties. For instance, graphene is a zero--bandgap semiconductor, while phosphorene has a $2$ $\mathrm{eV}$ natural bandgap. Also, the bandgaps observed in most TMDCs are larger than $1$ $\mathrm{eV}$. Therefore, different materials in the 2D domain present diverse electronic and optoelectronic applications.  

\begin{figure*}
   \centering   
   \includegraphics[scale=0.5]{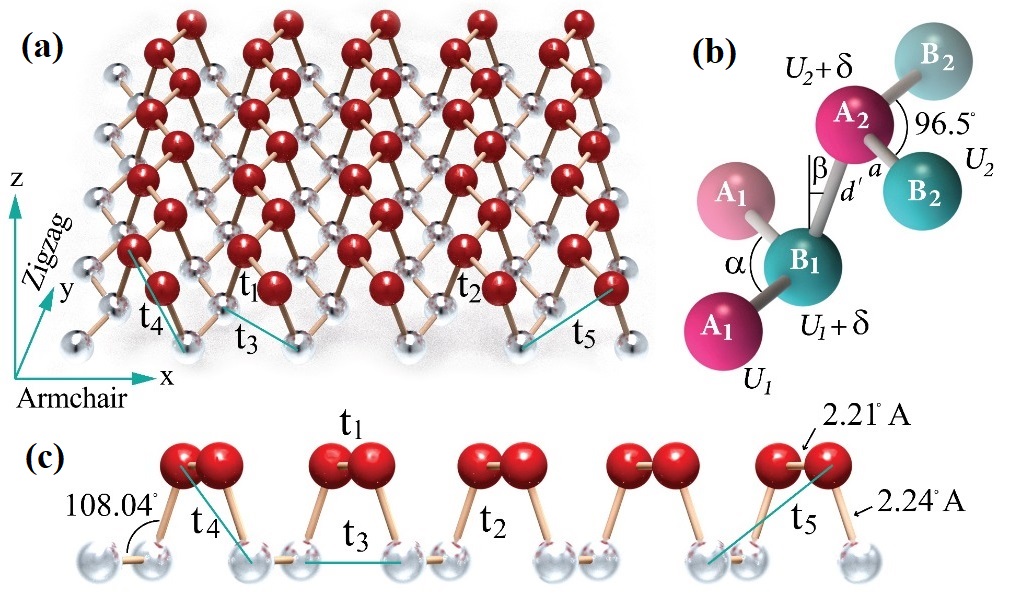}
   \caption{(a) Phosphorene structure and associated coupling parameters $t_i$ ($i=1,2,...,5$). The $x$ is armchair and $y$ is zigzag direction. (b) The phosphorene unit cell. The side view of four atoms in the phosphorene unit cell. The $A_1$ and $B_1$ atoms are positioned on the lower layer while $A_2$ and $B_2$ are situated on the upper layer. The intralayer distance between the atoms in a single unit cell is $a$ and the distance between $B_1$ and $A_2$ atoms in different layers is $d^\prime$. The parameters $U_1$, $U_2$, $\delta$ specify the various on--site energies. (c) The side view of phosphorene, including the coupling parameters, atomic distances, and the $\widehat{A_2 B_1 A_1}$ angle \citep{chaves2017theoretical}.} 
    \label{fig1}
\end{figure*}

In this study, we investigate phosphorene, the monolayer of black--phosphorus (Fig.~\ref{fig1}(a)). Several motivations exist for studying phosphorene’s response to THz laser radiation. First, in phosphorene, carrier mobility is much higher than other 2D materials such as TMDCs ($\leq 200$ $\mathrm{cm^2 V^{-1} s^{-1}}$). Also, phosphorene is comparable to graphene in its electron conduction rate ($650$ $\mathrm{cm^2 V^{-1} s^{-1}}$ at room temperature). This characteristic makes phosphorene a desirable material for high--frequency electronics \citep{dhanabalan2017emerging, viti2015black, xia2014rediscovering}. Second, though the large dark currents that dominate under a non--zero bias operation restrict graphene’s performance, phosphorene’s direct bandgap makes it a suitable choice in high--frequency optoelectronics, ultra--fast photonics, transparent photovoltaics, and photodetection \citep{viti2015black, gan2013chip, koppens2014photodetectors, luo2015microfiber, sun2016optical}. Undeniably, phosphorene’s $I_{on}/I_{off}$ ratio of $~10^5$ make it an appropriate choice for detecting THz frequency radiation \citep{das2014tunable, viti2015black}. Third, every phosphorus atom in phosphorene has a covalent bond with three adjacent phosphorus atoms. Hence, each of the $p$ orbitals holds a single pair of electrons. Unlike graphene, $sp^3$ hybridization prevents phosphorene from forming an atomically flat sheet.  The latter characteristic produces an intrinsic in--plane anisotropy, which deduces to an explicitly angle-dependent conductivity \citep{viti2015black}. Phosphorene’s puckered structure deduces to a strong anisotropy in electrical conductivity, and it is important to have novel devices with anisotropic properties. \citep{sun2016optical}. This is while the photonic and electronic properties of graphene and TMDCs are largely isotropic and do not exhibit a significant directional dependence \citep{xia2014rediscovering, peruzzini2019perspective}. Fourth, phosphorene is an inorganic material easily integrated with photonic or optoelectronic 2D materials, such as graphene or silicon--based materials \citep{viti2015black}. For these reasons, we selected phosphorene to investigate the linear and nonlinear responses of a 2D anisotropic material subjected to THz laser radiation. Ultrafast signal processing largely relies on nonlinear optics in which the higher power of the electric field determines the strength of the response. Although recent studies have studied graphene in the context of the linear and nonlinear transport effects in electric fields \citep{kheirabadi2018cyclotron, kheirabadi2016magnetic, drexler2013magnetic, glazov2014high, olbrich2016terahertz}, the linear and nonlinear current response in phosphorene has seldom been investigated \citep{kheirabadi2020magnetic}. This study aims to examine the current response of phosphorene subjected to a steady magnetic field. The present study results offer a new perspective on the anisotropic nonlinear optical properties of an anisotropic 2D material with possible applications in polarized optics, optical switching, and photodetection. The latter relies on the conversion of absorbed photons into an electrical signal that is probably the most explored black--phosphorus based photonic device \citep{guo2016black}. The calculations provided in this study are valid for $\hbar \omega \leq \epsilon_f$ where $\omega$ is the angular frequency of laser radiation, and $\epsilon_f$ is the Fermi level; semi--classical regime. Therefore, the study’s results are suitable for application in THz-- and microwave--  optoelectronic devices.

In this study, we demonstrate that subjected to a steady tilted magnetic field and THz laser radiation, a first-order AC current, a second-order DC current (ratchet effect), and a second-order AC current (second harmonic generation (SHG)) is deduced in phosphorene. The ratchet effect is a nonlinear response to a driving light field that results in a DC current. SHG is a second-order in electric field effect by which an AC current with two times the frequency related to the incoming AC radiation is generated. Also, due to an existing perpendicular magnetic field, the cyclotron resonance impacts the deduced currents’ strength in phosphorene. The cyclotron resonance of the magnetic ratchet effect and SHG have been investigated in isotropic materials such as 2D electron gas \citep{budkin2016ratchet}, and bilayer graphene \citep{kheirabadi2018cyclotron}. The effect is yet to be investigated in anisotropic materials, such as phosphorene, which we will examine in this study.
\section{Boltzmann equation}
We derived the tight--binding Hamiltonian of phosphorene under the effect of an in--plane magnetic field \citep{kheirabadi2020magnetic}.
The Hamiltonian of phosphorene in a steady in--plane magnetic field and in the basis of $(A_1,B_1,A_2,B_2)^{T}$ (Fig.~\ref{fig1}(b)) is
\begin{eqnarray}\label{HBP1}
H=\begin{pmatrix}
U_1 & f_1+f_3  & f_4  & f_2+f_5 \\ 
{f_1}^*+{f_3}^* & U_1 + \delta  & {f_2}^*+{f_5}^* & f_4\\ 
{f_4}^* & f_2+f_5 & U_2 + \delta  & f_1^{\prime}+f_3^{\prime} \\ 
{f_2}^*+{f_5}^* & {f_4}^*& {f_1^{\prime}}^*+{f_3^{\prime}}^*  &U_2 
\end{pmatrix}.\nonumber
\end{eqnarray}
We assume that $\textbf{b}=e d \textbf{B}_{\parallel} / 2 $ where $-e$ is the electron charge, $d$ is the interlayer distance, $\textbf{B}_{\parallel}$ is the in--plane magnetic field, in--plane momentum is $\mathbf{p}=(p_x,p_y,0)$, different on-site energies are $U_1$ and $U_2$, $\delta$ is the interlayer potential asymmetry, and $a_x$ and $a_y$ are the length of the unit cell into the $x$ (armchair) and $y$ (zigzag) directions, respectively (Fig.~\ref{fig1}(a and b)). Hence, we have
\begin{eqnarray}\label{f}
f_1 && = 2 t_1 \cos{\frac{a_y(p_y+b_x)}{2 \hbar}}\mathrm{exp}{\big[\frac{i(p_x-b_y )}{\hbar}a \cos{\frac{\alpha}{2}}}\big],\nonumber\\
f_1^\prime && = 2 t_1 \cos{\frac{a_y(p_y-b_x)}{2 \hbar}}\mathrm{exp}{\big[\frac{i(p_x+b_y )}{\hbar}a \cos{\frac{\alpha}{2}}}\big],\nonumber\\
f_2 && = t_2 \mathrm{exp}[-\frac{i}{\hbar}p_x d^\prime \sin\beta],\nonumber\\
f_3 && = 2 t_3 \cos{\frac{a_y(p_y+b_x)}{2 \hbar}}\nonumber\\
&&\quad \times \mathrm{exp}{\big[-\frac{i(p_x-b_y )}{\hbar}(2 d ^ \prime \sin \beta + a \cos{\frac{\alpha}{2})}}\big],\nonumber\\
f_3^\prime && = 2 t_3 \cos{\frac{a_y(p_y-b_x)}{2 \hbar}}\nonumber\\
&&\quad \times \mathrm{exp}{\big[-\frac{i(p_x+b_y )}{\hbar}(2 d ^ \prime \sin \beta + a \cos{\frac{\alpha}{2})}}\big],\nonumber\\
f_4 && = 4 t_4 \cos[\frac{p_x}{\hbar}(d^\prime \sin\beta+a \cos\frac{\alpha}{2})] \cos[\frac{p_y}{\hbar}a\sin\frac{\alpha}{2}],\nonumber\\
f_5 && = t_5 \mathrm{exp}[i \frac{p_x}{\hbar}(a_x-d^\prime \sin\beta)].\nonumber
\end{eqnarray}
In the above-mentioned equations, the intralayer coupling is $t_1$, the interlayer couplings are $t_2$, $t_4$ and $t_5$, $d^{\prime}$ is the distance between $B_1$ and $A_2$ atoms, $\alpha$ is $\widehat{A_1B_1A_1}=\widehat{B_2A_2B_2}= 96.5 ^ \circ$, $\beta$ is $\widehat{A_2B_1A_1}-90^\circ=108.04^{\circ}-90^\circ=18.04 ^ \circ$ \citep{chaves2017theoretical, pereira2015landau} (Fig.~\ref{fig1}(c)), the upper layer is located at $d/2$, and the lower layer is located at $-d/2$ (Fig.~\ref{fig1}). Based on this Hamiltonian and according to the perturbation theory, we can derive the perturbed valence and conduction bands \citep{kheirabadi2020magnetic, kheirabadi2016magnetic}. By this assumption, we have considered the orbital effect of an in--plane magnetic field on the electrons. Furthermore, we assume that the redistribution of charge carriers in the momentum and energy space is induced by an in--plane laser radiation incident on the phosphorene plane. Consequently, we assume that the applied AC electric field of THz laser radiation is 
\begin{equation}\label{E}
\mathbf{E}_{\parallel}(t)=\mathbf{E}_{\parallel}\mathrm{exp}(-i \omega t)+\mathbf{E}_{\parallel}^* \mathrm{exp}(i \omega t),\nonumber
\end{equation}
where $\mathbf{E}_{\parallel}=(E_x,E_y)$, and $\mathbf{E}_{\parallel}^*=(E_x^*,E_y^*)$.

Out of an equilibrium condition caused by the laser radiation, the electron distribution function is dependent on the momentum $\mathbf{p}$ and time $t$; $f(\mathbf{p},t)$. If we consider that the system under study is spatially homogeneous, $f(\mathbf{p},t)$ satisfies the following Boltzmann kinetic equation
\begin{eqnarray}\label{be}
-e\bigg( \mathbf{E}_\parallel + \mathbf{V}_g \times \mathbf{B}_\perp\bigg) \cdot \mathbf{\triangledown}_p f(\mathbf{p},t)+\frac{\partial f(\mathbf{p},t)}{\partial t}=S\{ f\}\ ,\nonumber\\
\end{eqnarray}
where the group velocity is $\mathbf{V}_g=V_{g,x}\cos \phi \mathbf{\hat{i}}+V_{g,y}\sin \phi \mathbf{\hat{j}}$, $\phi$ is the polar angle of momentum, $\mathbf{B}_\perp$ is the perpendicular magnetic field, and $S\{f\}$ is the collision integral. Also, $V_{g,x}=p/m_{xx}$ and $V_{g,y}=p/m_{yy}$, where $p={\left | \textbf{p} \right |}$, $m_{xx}$ and $m_{yy}$ are effective mass along the $x$ and $y$ directions, respectively \citep{lowrelax}. To solve the Boltzmann equation, Eq.~\ref{be}, we consider the distribution function as the following series
\begin{equation}\label{fmn}
f(\mathbf{p},t)=\sum_{n,m}f_m^n e^{i m \phi-i n \omega t}, \nonumber
\end{equation} 
where $f_m^n$ coefficients are functions of the total energy of an electron, $\epsilon$, and m and n are integers. Then, we multiply the Boltzmann equation in $\mathrm{exp}(-i j \phi+i l \omega t)$, where $j$ and $l$ are integers and integrate over a period of $2 \pi$ of angle $\phi$ and a period of time; $t$. Therefore, coupled equations between different $f_m^n$ coefficients are achieved in the following form
\begin{eqnarray}\label{fjl}
\gamma^{l,j} f_{j}^{l}&&=\alpha _{j-1}f_{j-1}^{l-1}+\widetilde{\alpha}_{j-1}f_{j-1}^{l+1}+ \beta_{j+2} f^l_{j+2}\nonumber\\
&&+\widetilde{\beta}_{j-2}f^l_{j-2}+ \eta _{j+1}f_{j+1}^{l-1}+\widetilde{\eta}_{j+1}f_{j+1}^{l+1}+\delta S_{j}^{l}.
\end{eqnarray}

In Eq.~\ref{fjl}, $\gamma^{l,j}=\tau _{\left | j \right |,\mathbf{p}}^{-1}- i l\omega+i j \omega_{c_{+}}$, where $\tau _{\left | j \right |,\mathbf{p}}^{-1}$ is the relaxation time of the jth angular harmonic of the electron distribution function, $\omega_{c_{+}}=e B_\perp /(2 m_{+})$ is the cyclotron frequency including the cyclotron resonance effect, and $1/m_{+}=1/m_{xx} + 1/ m_{yy}$. Also, linear in electric field $\alpha$, $\beta$, and $\eta$ operators are
\begin{eqnarray}
\alpha_j&&= \frac{e (E_x-iE_y)}{2} \left(-\frac{j}{p} +\frac{\partial }{\partial p} \right ),\nonumber\\
\tilde{\alpha}_j&&= \frac{e (E_{x}^{*}-iE_{y}^{*})}{2} \left(-\frac{j}{p} +\frac{\partial }{\partial p} \right ),\nonumber\\
\beta_j&&=\frac{1 }{2}\omega_{c_{-}}\big(i p \frac{\partial }{\partial p}+j\big),\nonumber\\
\widetilde{\beta}_{j}&&=\frac{1 }{2}\omega_{c_{-}}\big(-i p \frac{\partial }{\partial p}+j\big),\nonumber\\
\eta_j && = \frac{e (E_x+iE_y)}{2} \left (\frac{j}{p} +\frac{\partial }{\partial p}\right ),\nonumber\\
\tilde{\eta}_j && = \frac{e (E_{x}^{*}+iE_{y}^{*})}{2} \left( \frac{j}{p} +\frac{\partial }{\partial p} \right ).\nonumber
\end{eqnarray}
Here, we have $\omega_{c_{-}}=e B_{\perp}/(2 m_{-})$ and $1/m_{-}=1/m_{xx}-1/m_{yy}$. In Eq.~\ref{fjl}, $\delta S_j^l$ is the correction to the scattering caused by an in--plane magnetic field.
 
For an anisotropic 2D electron gas such as phosphorene, where $\xi$ is the unit matrix of the electric field, the relaxation time of the jth angular harmonic, $\tau _{\left | j \right |,\mathbf{p}}^{-1}$, is \citep{relaxtime, lowrelax}
\begin{eqnarray}\label{tau}
\tau^{-1}_{\left | j \right |, \mathbf {p}}(\mathbf \xi, \mathbf p )&&=\frac{2\pi}{\hbar}\sum_{\mathbf p^{\prime}}  \left| \langle \mathbf{p^{\prime}}\left| \delta H \right|\mathbf{p}\rangle\right|^2 \delta \left(\epsilon_p-\epsilon_{p^{\prime}} \right)\nonumber\\
&& \quad \times \left\{1-\frac{ [ \mathbf{\xi} . \mathbf{V}_g(\mathbf{p^{\prime}}) ] \tau_{\left | j \right |, \mathbf {p^{\prime}}}}{[ \mathbf{\xi} . \mathbf {V}_g(\mathbf{p}) ] \tau_{\left | j \right |, \mathbf {p}}} \right\}.
\end{eqnarray} 
Note that for isotropic materials, we have $\omega_{c_{-}}=0$ and $\omega_{c_{+}}= e B V_g / p$ and $\tau_{\left | j \right |, \mathbf {p}}(\mathbf \xi, \mathbf p )= \tau_{\left | j \right |}$. So, Eq.~\ref{fjl} and \ref{tau} are also valid for isotropic materials \citep{budkin2016ratchet, olbrich2013giant, dantscher2015cyclotron, kheirabadi2018cyclotron}.

For static impurities, we can show that \citep{kheirabadi2016magnetic, kheirabadi2018cyclotron,kheirabadi2020magnetic}
\begin{eqnarray}\label{deltah}
\delta H=\sum_{j=1}^{N_{imp}}\hat{Y}u(\mathbf{r}-\mathbf{R_j}),
\end{eqnarray}
where $N_{imp}$ is the number of impurities, $u(\mathbf{r}-\mathbf{R_j})$ describes the spatial dependence of the impurity potential and $\hat{Y}$ is a dimensionless matrix describing the additional degree of freedom related to the structure within the unit cell. We neglect the interference between different impurities and employ the Fourier transform of the impurity potential and perform a harmonic expansion of the impurity potential; so, we have
\begin{eqnarray}
\lvert \tilde {u}(\mathbf{p^{\prime}-p}) \rvert ^2 =\sum_{m^{\prime}}\nu_{m^{\prime}}e^{im^{\prime}(\phi^{\prime}-\phi)}. \nonumber
\end{eqnarray}
In addition, we assume that the electrons are trapped in a huge box with the length $L$ and under a periodic potential. Hence, the current density is
\begin{eqnarray}\label{current}
\mathbf{J}=-\frac{g}{L^2}\sum_{\mathbf{p}}e \mathbf{V}_{g} f(\mathbf {p}, t),\nonumber
\end{eqnarray} 
where $g$ is the spin degeneracy factor $(g=2)$. In thermal equilibrium, and wherever there is no electric field, only the $f_0^0$ harmonic generated by the Fermi--Dirac distribution function is nonzero. However, when an AC electric field is applied to the system, other harmonics arise, and this gives the possibility of having a linear (first-order) or nonlinear (higher orders) in electric field currents. In this study, we consider all nonzero first order or second order in electric field currents. Hence, based on coupled equations (Eq.~\ref{fjl}), $f_j^l$ harmonics in terms of the equilibrium distribution function ($f_0^0$) should be calculated. In addition, we assume that the system is a degenerate electron gas at low temperature condition; so, we have $\partial f_0^{0}/ \partial \epsilon \approx-\delta(\epsilon-\epsilon_f)$.             
\section{Phosphorene}
To find $f_j^l$ harmonics in Eq.~\ref{fjl}, first, we should find the correction to the scattering caused by the in--plane magnetic field, $\delta S_j^l$. We have 
\begin{eqnarray}\label{deltas}
\delta S _j^l&&=L^2 \int_{0}^{\infty}\bigg[ \sum_m f_m^l \Gamma(\epsilon) d\epsilon \nonumber\\
&& \times \int_{0}^{2 \pi}\delta W_{\mathbf{p^{\prime}}\mathbf{p}}\bigg( e^{im\phi^{\prime}-ij\phi}-e^{im\phi-ij\phi}\bigg) \frac{d \phi}{2 \pi} \frac{d \phi^{\prime}}{2 \pi}\bigg],\nonumber\\
\end{eqnarray} 
where $\Gamma(\epsilon)$ is the electronic density of states per spin and per unit area and $\delta W_{\mathbf{p^{\prime}}\mathbf{p}}$ is the change of the scattering rate caused by the in--plane magnetic field. The scattering rate of an electron passes through a phosphorene in a steady magnetic field to the linear order in $\mathbf{B}_\parallel$ and momentum is calculated before \citep{kheirabadi2020magnetic}. Accordingly, for the case of the asymmetric disorder where the lower layer ($\zeta=1$) or the upper layer ($\zeta=-1$) symmetry is broken by the disorder ($z\rightarrow-z$ asymmetry), $\hat{Y}$ in Eq.~\ref{deltah} is equal to $\big( \hat{I}+\zeta \hat{\sigma_z}\otimes\hat{I} \big)/2$, where $\hat{I}$ is the $2 \times 2$ unit matrix, and $\sigma_z$ is a Pauli matrix. Consequently, $\delta W_{\mathbf{p^{\prime}}\mathbf{p}}$ has the following general form:
\begin{eqnarray}\label{deltaw}
\delta W_{\mathbf{p^{\prime}} \mathbf{p}} (\zeta,U_1,U_2,\delta) && = \frac{2 \pi}{\hbar} \frac{n_{imp}}{L^2}  \lvert \tilde{u}( \bf{p}^{\prime}- \bf{p} ) \rvert ^ 2 \delta(\epsilon_{p^\prime}-\epsilon_p)\nonumber\\
 && \times \bigg\{\frac{1}{\hbar^2} C_1(\zeta,U_1,U_2,\delta) b_y p (\cos \phi + \cos\phi^ \prime)\nonumber\\
&& + \frac{1}{\hbar^2} C_2 (\zeta,U_1,U_2,\delta) b_x p ( \sin \phi +\sin \phi^ \prime ) \bigg\}.\nonumber\\ 
\end{eqnarray}
In Eq.~\ref{deltaw}, the $n_{imp}=N_{imp}/L^2$, $C$ coefficients are dependent on on--site energies and disorder types. Consequently, according to Eq.~\ref{deltas}, we can show that 
\begin{allowdisplaybreaks}
\begin{eqnarray}
\delta S_{0}^{l}&&=0, \nonumber\\
\delta S_{1}^{l}&&=\Lambda  \big( C_1 B_y +i C_2 B_x\big) f_2^l,\nonumber\\
\delta S_{-1}^{l}&&=\Lambda  \big( C_1 B_y -i C_2 B_x \big) f_{-2}^l,\nonumber\\
\delta S_{2}^{l}&&=\Lambda  \big( C_1 B_y - i C_2 B_x\big) f_1^l,\nonumber\\
\delta S_{-2}^{l}&&=\Lambda  \big(C_1 B_y + i C_2 B_x\big) f_{-1}^l,\nonumber
\end{eqnarray}
\end{allowdisplaybreaks}
where $\Lambda = e d \pi n_{imp}\Omega \Gamma(\epsilon) p /2 \hbar^3$ and $\Omega = -(\nu_0-\nu_2)$.
\section{First order AC current}
The first order AC current is the result of $f_1^1$ and $f_{-1}^1$ harmonics and the complex conjugate of those terms; $f_m^n=\big( f_{-m}^{-n}\big)^*$. We can show that the first order AC current is 
\begin{equation}
\mathbf{J}=2 Re \{\sigma \mathbf{E} e^{-i \omega t}\}, \nonumber
\end{equation}
where $\sigma$ is the conductivity tensor. The conductivity tensor components are $\sigma_{ii}=\sigma_{0i} \sigma^{\prime}_{ii}$ and $\sigma_{ij}=\sigma_{0i} \sigma^{\prime}_{ij}$, where $i$ and $j$ indicate the $x$ or $y$ directions and
\begin{eqnarray}\label{sigma0i}
\sigma_{0i}&&=\frac{g e^2}{2} \Gamma(\epsilon) C_{ph} p V_{g,i},
\end{eqnarray}
\begin{eqnarray}\label{sigmaxx}
\sigma^{\prime}_{xx}=\sigma^{\prime}_{yy}&&= \frac{(1-i \omega \tau _{1,\mathbf{p}})\tau _{1,\mathbf{p}} }{(1-i \omega \tau _{1,\mathbf{p}} )^2+(\omega_{c_{+}} \tau _{1,\mathbf{p}})^2},
\end{eqnarray}
\begin{eqnarray}\label{sigmaxy}
\sigma^{\prime}_{xy}=- \frac{(\omega_{c_{+}}+\omega_{c_{-}}+i \omega_{c_{-}}/2) \tau^2 _{1,\mathbf{p}}}{(1-i \omega \tau _{1,\mathbf{p}} )^2+(\omega_{c_{+}} \tau _{1,\mathbf{p}})^2}
\end{eqnarray}
\begin{eqnarray}\label{sigmayx}
\sigma^{\prime}_{yx}&&= \frac{(\omega_{c_{+}}-\omega_{c_{-}}-i \omega_{c_{-}}/2) \tau^2 _{1,\mathbf{p}}}{(1-i \omega \tau _{1,\mathbf{p}} )^2+(\omega_{c_{+}} \tau _{1,\mathbf{p}})^2}
\end{eqnarray}
Here, we have $\omega_{c_{+}}+\omega_{c_{-}}= e B_{\perp}/ m_{xx}$ and $\omega_{c_{+}}-\omega_{c_{-}}=eB_{\perp}/m_{yy}$. These calculations are valid for a degenerate electron gas, $\epsilon _f \gg k_B T$. All the parameters are evaluated on the Fermi surface and the above--mentioned results are in agreement with those related to isotropic materials \citep{kheirabadi2018cyclotron}. In Eq.~\ref{sigma0i}, we have
\begin{eqnarray}
\frac{\partial }{\partial p}&&=C_{ph}p \frac{\partial }{\partial \epsilon},\nonumber\\
C_{ph}&&=s\frac{2}{\hbar^2}\bigg[ \frac{\gamma^2}{E_g}+\big(\eta_{v/c}+\nu_{v/c}\big)\bigg],\nonumber
\end{eqnarray}
where $s$ is the band index, and it is $+1$ for the conduction band and $-1$ for the valence band. $E_g$ is the direct energy gap, $E_g=0.912$ $\mathrm{eV}$, $\gamma=0.480$ $\mathrm{eV nm}$, $\eta_v=0.038$ $\mathrm{eV nm^{2}}$, $\nu_v=0.030$ $\mathrm{eV nm^2}$, $\eta_c= 0.008$ $\mathrm{eV nm^{2}}$, and $\nu_c=0.030$ $\mathrm{eV nm^2}$ are from Ref.~\citep{AsgariHamil}.  

According to Eqs.~\ref{sigmaxx} to \ref{sigmayx}, there are resonances at $\omega=\pm \omega_{c_{+}}$. At these frequencies, the current is much larger than at $\omega_{c_{+}}=0$.
\section{Ratchet effect}
The ratchet effect is a nonlinear response to a driving light field where $f_1^0$ and $f_{-1}^0$ harmonics result in a DC current. The ratchet current is calculated based on the following equation:
\begin{equation}
\mathbf{J}=-\frac{ge}{L^2}\sum_\mathbf{p} \mathbf{V}_g \big( f_1^0 e^ {i \phi}+ f_{-1}^0 e^ {- i \phi}\big).\nonumber
\end{equation}
For $\Theta_1=\lvert E_x \rvert ^2- \lvert E_y \rvert^2$, $\Theta_2=E_x E_y^*+E_x^* E_y$, and $\Theta_3=i \big( E_x E_y^*-E_x^* E_y \big)$, we can show that the current is
\begin{eqnarray}\label{jxlr}
J_x&&=\lvert {E} \rvert^2 \big( B_y ^{\prime} Re [M_{2,x}]+ B_x^{\prime} Im [M_{2,x}]\big)\nonumber\\
&&+\Theta_1\big( B_y ^{\prime} Re [M_{1,x}]- B_x^{\prime} Im [M_{1,x}]\big)\nonumber\\
&&+\Theta_2 \big( B_y ^{\prime} Im [M_{1,x}]+ B_x^{\prime} Re [M_{1,x}]\big)\nonumber\\
&&+\Theta_3\big( B_y ^{\prime} Re [M_{3,x}]+ B_x^{\prime} Im [M_{3,x}]\big),\nonumber
\end{eqnarray}
\begin{eqnarray}\label{jylr}
J_y&&=\lvert{E} \rvert ^2 \big(-B_y ^{\prime} Im [M_{2,y}]+ B_x^{\prime} Re [M_{2,y}]\big)\nonumber\\
&&- \Theta_1\big( B_y ^{\prime} Im [M_{1,y}]+ B_x^{\prime} Re [M_{1,y}]\big)\nonumber\\
&&+\Theta_2 \big( B_y ^{\prime} Re [M_{1,y}] - B_x^{\prime} Im [M_{1,y}]\big)\nonumber\\
&&+\Theta_3\big(- B_y ^{\prime} Im [M_{3,y}]+ B_x^{\prime} Re [M_{3,y}]\big),\nonumber
\end{eqnarray}
where $B_x ^{\prime}=C_2 B_x$ and $B_y ^{\prime}=C_1 B_y$. In addition, the $M$ coefficients are
\begin{eqnarray}
M_{1,i}&&=-\frac{g e^3}{4} C_{ph}p \big( \frac{1}{\gamma^{-1,1}}+ \frac{1}{\gamma^{1,1}} \big)\nonumber\\
&& \times \bigg\{ \frac{V_{g,i} \Lambda}{\gamma^{0,1}\gamma^{0,2} p }\Gamma(\epsilon) + C_{ph} ( \Gamma(\epsilon)  \frac{V_{g,i} \Lambda}{\gamma^{0,1}\gamma^{0,2} } p \big)^{\prime} \bigg\},\nonumber
\end{eqnarray} 
\begin{eqnarray}
M_{2,i}&& =\frac{g e^3}{4}C_{ph}p \Lambda \big( \frac{1}{\gamma^{-1,2}\gamma^{-1,1}}+ \frac{1}{\gamma^{1,2}\gamma^{1,1}} \big)\nonumber\\
&& \times\bigg\{ \frac{2 V_{g,i}}{\gamma^{0,1} p }\Gamma(\epsilon)
 - C_{ph} ( \Gamma(\epsilon)  \frac{V_{g,i}}{\gamma^{0,1} } p \big)^{\prime} \bigg\},\nonumber
\end{eqnarray} 
\begin{eqnarray}
M_{3,i}&&=\frac{g e^3}{4}C_{ph}p \Lambda \big( \frac{1}{\gamma^{1,2}\gamma^{1,1}}- \frac{1}{\gamma^{-1,2}\gamma^{-1,1}} \big)\nonumber\\
&& \times  \bigg\{ \frac{2 V_{g,i}}{\gamma^{0,1} p }\Gamma(\epsilon)
 - C_{ph} ( \Gamma(\epsilon)  \frac{V_{g,i}}{\gamma^{0,1} } p \big)^{\prime} \bigg\},\nonumber
\end{eqnarray}
where $(\ldots)^{\prime} \equiv \partial (\ldots)/ \partial \epsilon$ and all parameters are evaluated on the Fermi surface. Also, for $\omega_{c_{+}}=0$, where the perpendicular magnetic field is zero, the above results are in agreement with previous results \citep{kheirabadi2020magnetic}.
 
\begin{figure}[h!]
   \centering   
   \includegraphics[scale=0.6]{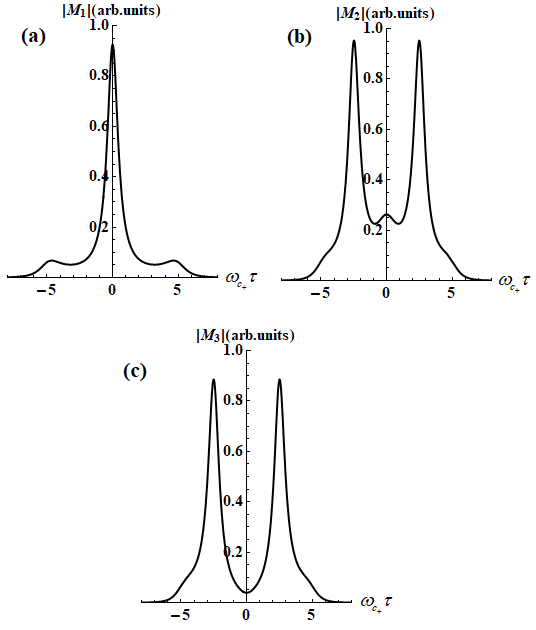}
   \caption{The $\omega_{c_{+}} \tau$ dependent magnitude of $M$ coefficients considering a momentum-independent scattering time $\tau_1=\tau_2=\tau$ and $\omega\tau=5$.} 
    \label{fig2}
\end{figure}

Moreover, $M_{1,i}$ is the response to the linearly polarized light, $M_{2,i}$ is the response to the unpolarized light, and $M_{3,i}$ is the response to the circularly polarized light. For linearly polarized light, we can assume that $E_x^{*}=E_x=(E_0/2) \cos{\theta}$ and $E_y^{*}=E_y=(E_0/2) \sin{\theta}$, where $\theta$ is the polarization angle. We also assume that $B^{\prime}_\parallel=\big({B^{\prime}_x}^2+{B^{\prime}_y}^2\big)^{1/2}$ and $\varphi^{\prime}=\arctan\big( B^{\prime}_y / B^{\prime}_x \big)$. For linearly polarized light, we can show that the currents in the $x$ and $y$ directions are
\begin{eqnarray}
J_x&&=\frac{E_0^2}{4} B_\parallel^{\prime} \bigg\{\left\lvert M_{2,x} \right\rvert \sin(\varphi^{\prime}+\chi_{2,x})\nonumber\\
&& \quad \quad \quad \quad +\left\lvert M_{1,x} \right\rvert \cos(2 \theta+\varphi^\prime-\chi_{1,x}+\frac{\pi}{2})\bigg\},\label{jxlr}\\
J_y&&=\frac{E_0^2}{4} B_\parallel^{\prime} \bigg\{\left\lvert M_{2,y} \right\rvert \cos(\varphi^{\prime}+\chi_{2,y})\nonumber\\
&&\quad \quad \quad \quad +\left\lvert M_{1,y} \right\rvert \sin(2 \theta+\varphi^\prime-\chi_{1,y}+\frac{\pi}{2})\bigg\},\label{jylr}
\end{eqnarray}
where $\chi_{1,i}=arg(M_{1,i})$ and $\chi_{2,i}=arg(M_{2,i})$. For unpolarized light, $M_{1,i}$ related terms are equal to zero, but $M_{2,i}$ related terms survive. In addition, $M_{1,i}$ related current has a notable resonance for $\omega=\pm \omega_{c_{+}}$ and the strength of the ratchet effect is highest for $\omega_{c_{+}}=0$ (Fig.~\ref{fig2}). Current induced by unpolarized light, $M_{2,i}$ has resonances for $\omega=\pm 2 \omega_{c_{+}}$ that deduces to a DC current that is much larger in comparison to where $\omega_{c_{+}}=0$; $B_\perp=0$. The other resonance is related to $\omega=\pm \omega_{c_{+}}$. However, in this case, the deduced DC current is smaller than where $\omega_{c_{+}}=0$.   

For circularly polarized light, we can show that $E_x^{*}=E_x=E_0/2$ and $E_y^{*}=-E_y=- i \mu E_0/2$, where $\mu=1(-1)$ for the left- (right-) handed circularly polarized light. For this radiation type, where $\chi_{3,i}=arg(M_{3,i})$, we can show that the current density is
\begin{eqnarray}
J_x&&=\frac{E_0^2}{2} B_\parallel^{\prime} \bigg\{\left\lvert M_{2,x} \right\rvert \sin(\varphi^{\prime}+\chi_{2,x})\nonumber\\
&&\quad \quad \quad \quad + \mu \left\lvert M_{3,x} \right\rvert \sin(\varphi^\prime+\chi_{3,x})\bigg\},\label{jxcr}\\
J_y&&=\frac{E_0^2}{2} B_\parallel^{\prime} \bigg\{\left\lvert M_{2,y} \right\rvert \cos(\varphi^{\prime}+\chi_{2,y})\nonumber\\
&&\quad \quad \quad \quad + \mu \left\lvert M_{3,y} \right\rvert \cos(\varphi^\prime+\chi_{3,y})\bigg\}.\label{jycr}
\end{eqnarray}
The above equations show that the response to circularly polarized light is dependent on the radiation helicity. For $M_{3,i}$, the major resonance effects are related to $\omega=\pm 2 \omega_{c_{+}}$. There is also a resonance effect for $\omega=\pm \omega_{c_{+}}$ and both effects cause a current that is stronger than the current where $B_\perp=0$ (Fig.~\ref{fig2}). According to Eqs.~\ref{jxlr} to \ref{jycr}, the direction of the ratchet current for circularly and linearly polarized lights is dependent on the $\varphi^{\prime}$ and $\chi$ phases. 
\section{SHG effect}
Assuming $\Theta_4=E_x^2-E_y^2$, and $\Theta_5=2 E_x E_y$ and $\Theta_6=E_x^2+E_y^2$, we can show that the general current form for a degenerate electron gas is
\begin{eqnarray}
J_x &&=2 Re \bigg\{ \big[ \Theta_4 \big( N_{1,x} B_y^{\prime}+N_{2,x} B_x^{\prime} \big)\nonumber\\
&& \quad \quad \quad \quad + \Theta_5 \big( -N_{2,x} B_y^{\prime}+N_{1,x} B_x^{\prime} \big)\nonumber\\
&&\quad \quad \quad \quad + \Theta_6 \big(N_{3,x} B_y^{\prime}+N_{4,x} B_x^{\prime} \big) \big] e^{-2 i \omega t} \bigg\},\label{jx2}\\
J_y &&=2 Re \bigg\{ \big[ \Theta_4 \big( N_{2,y} B_y^{\prime} - N_{1,y} B_x^{\prime} \big) + \Theta_5 \big( N_{1,y} B_y^{\prime}+N_{2,y} B_x^{\prime} \big)\nonumber\\
&& \quad \quad \quad \quad + \Theta_6 \big(-N_{4,y} B_y^{\prime}+N_{3,y} B_x^{\prime} \big) \big] e^{-2 i \omega t} \bigg\},\label{jy2}
\end{eqnarray}
$N$ coefficients in Eqs.~\ref{jx2} and \ref{jy2} are
\begin{eqnarray}
N_{1,i}&&=-\frac{g e^3}{8}\big[ V_{g,i} C_{ph} \Lambda \Gamma(\epsilon) \big( \frac{1}{\gamma^{2,1}\gamma^{2,2}\gamma^{1,1}}+ \frac{1}{\gamma^{2,-1}\gamma^{2,-2}\gamma^{1,-1}}\big) \nonumber\\
&&\quad \quad \quad \quad+ C_{ph}^2 \frac{p}{\gamma^{1,1}}\big( \Gamma(\epsilon)\frac{V_{g,i} \Lambda}{\gamma^{2,1}\gamma^{2,2}} p \big)^{\prime} \nonumber\\
&&\quad \quad \quad \quad+ C_{ph}^2 \frac{p}{\gamma^{1,-1}}\big( \Gamma(\epsilon) \frac{V_{g,i} \Lambda}{\gamma^{2,-1}\gamma^{2,-2}} p \big)^{\prime} \big],\nonumber\\
N_{2,i}&&=-\frac{g e^3}{8}i \big[ V_{g,i} C_{ph} \Lambda \Gamma(\epsilon) \big( \frac{1}{\gamma^{2,1}\gamma^{2,2}\gamma^{1,1}} - \frac{1}{\gamma^{2,-1}\gamma^{2,-2}\gamma^{1,-1}}\big) \nonumber\\
&&\quad \quad \quad \quad+ C_{ph}^2 \frac{p}{\gamma^{1,1}}\big( \Gamma(\epsilon)\frac{V_{g,i} \Lambda}{\gamma^{2,1}\gamma^{2,2}} p \big)^{\prime}\nonumber\\
&&\quad \quad \quad \quad- C_{ph}^2 \frac{p}{\gamma^{1,-1}}\big( \Gamma(\epsilon) \frac{V_{g,i} \Lambda}{\gamma^{2,-1}\gamma^{2,-2}} p \big)^{\prime} \big],\nonumber\\
N_{3,i}&&=\frac{g e^3}{8} \big[2 V_{g,i}  C_{ph} \Lambda \Gamma(\epsilon) \big( \frac{1}{\gamma^{2,1}\gamma^{1,2}\gamma^{1,1}} + \frac{1}{\gamma^{2,-1}\gamma^{1,-2}\gamma^{1,-1}}\big) \nonumber\\
&&\quad \quad \quad \quad- C_{ph}^2 \frac{p \Lambda}{\gamma^{1,2}\gamma^{1,1}}\big( \Gamma(\epsilon)\frac{V_{g,i} }{\gamma^{2,1}} p \big)^{\prime}\nonumber\\
&&\quad \quad \quad \quad- C_{ph}^2 \frac{p \Lambda}{\gamma^{1,-2}\gamma^{1,-1}}\big( \Gamma(\epsilon)\frac{V_{g,i} }{\gamma^{2,-1}} p \big)^{\prime} \big],\nonumber\\
N_{4,i}&&=-\frac{i g e^3}{8} \big[2 V_{g,i}  C_{ph} \Lambda\Gamma(\epsilon) \times\nonumber\\
&&\quad \quad \quad \quad\big( \frac{1}{\gamma^{2,1}\gamma^{1,2}\gamma^{1,1}} - \frac{1}{\gamma^{2,-1}\gamma^{1,-2}\gamma^{1,-1}}\big) \nonumber\\
&&\quad \quad \quad \quad- C_{ph}^2 \frac{p \Lambda}{\gamma^{1,2}\gamma^{1,1}}\big( \Gamma(\epsilon)\frac{V_{g,i} }{\gamma^{2,1}} p \big)^{\prime}\nonumber\\
&&\quad \quad \quad \quad+ C_{ph}^2 \frac{p \Lambda}{\gamma^{1,-2}\gamma^{1,-1}}\big( \Gamma(\epsilon)\frac{V_{g,i} }{\gamma^{2,-1}} p \big)^{\prime} \big].\nonumber
\end{eqnarray}
Here, all parameters are evaluated on the Fermi surface. Note that, for zero perpendicular magnetic field, $\omega_{c_{+}}=0$, $N_{2,i}$ and $N_{4,i}$ vanish. Also, assuming $\psi_{1,i}=arg(N_{1,i})$, $\psi_{2,i}=arg(N_{2,i})$ and $\psi_{3,i}=arg(N_{3,i})$, we can show that the current density deduced by a linearly polarized light is
\begin{eqnarray}\label{jxsl}
J_x && =\frac{E_0^2}{2} B_\parallel^{\prime} \bigg\{\left\lvert N_{1,x} \right\rvert \sin(2 \theta+\varphi^{\prime}) \cos(2 \omega t - \psi_{1,x})\nonumber\\
&& \quad \quad \quad \quad +\left\lvert N_{2,x} \right\rvert \cos(2 \theta+\varphi^{\prime}) \cos(2 \omega t - \psi_{2,x})\nonumber\\
&& \quad \quad \quad \quad +\left\lvert N_{3,x} \right\rvert \sin\varphi^{\prime} \cos(2 \omega t - \psi_{3,x})\nonumber\\
&& \quad \quad \quad \quad +\left\lvert N_{4,x} \right\rvert \cos\varphi^{\prime} \cos(2 \omega t - \psi_{4,x})\bigg\},
\end{eqnarray}
\begin{eqnarray}\label{jysl}
J_y && =\frac{E_0^2}{2} B_\parallel^{\prime} \bigg\{- \left\lvert N_{1,y} \right\rvert \cos(2 \theta+\varphi^{\prime}) \cos(2 \omega t - \psi_{1,y})\nonumber\\
&& \quad \quad \quad \quad +\left\lvert N_{2,y} \right\rvert \sin(2 \theta+\varphi^{\prime}) \cos(2 \omega t - \psi_{2,y})\nonumber\\
&& \quad \quad \quad \quad +\left\lvert N_{3,y} \right\rvert \cos\varphi^{\prime} \cos(2 \omega t - \psi_{3,y})\nonumber\\
&& \quad \quad \quad \quad -\left\lvert N_{4,y} \right\rvert \sin\varphi^{\prime} \cos(2 \omega t - \psi_{4,y})\bigg\}.
\end{eqnarray}
Consequently, $\varphi^{\prime}$ affects the deduced current direction, and $\psi$ phases determine the time--lag between the incoming radiation and the deduced current. For unpolarized light, only $N_{3,i}$ and $N_{4,i}$ related currents survive.
 
\begin{figure}[h!]
   \centering   
   \includegraphics[scale=0.53]{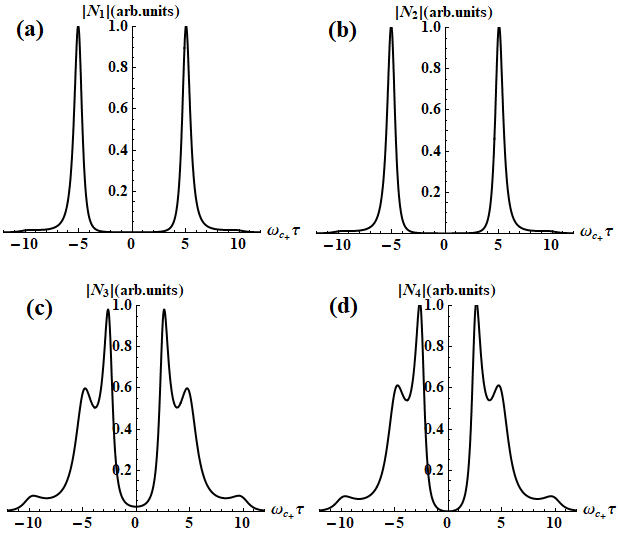}
   \caption{The $\omega_{c_{+}} \tau$ dependent magnitude of $N$ coefficients considering a momentum-independent scattering time $\tau_1=\tau_2=\tau$ and $\omega\tau=5$.} 
    \label{fig3}
\end{figure}

For circularly polarized light, the current density is 
\begin{eqnarray}
J_x &&=E_0^2 B_\parallel^{\prime} \bigg\{\left\lvert N_{1,x} \right\rvert  \sin(2 \omega t - \psi_{1,x}+\mu \varphi^{\prime})\nonumber\\
&& \quad \quad \quad \quad + \left\lvert N_{2,x} \right\rvert  \cos(2 \omega t - \psi_{2,x}+\mu \varphi^{\prime})\bigg\},\nonumber\\
J_y&& =E_0^2 B_\parallel^{\prime} \bigg\{-\left\lvert N_{1,y} \right\rvert  \cos(2 \omega t - \psi_{1,y}+\mu \varphi^{\prime})\nonumber\\
&& \quad \quad \quad \quad + \mu \left\lvert N_{2,y} \right\rvert  \sin(2 \omega t - \psi_{2,y}+\mu \varphi^{\prime})\bigg\}.\nonumber
\end{eqnarray}
Accordingly, $\varphi^{\prime}$ and $\psi$ phases determine the time--lag between the incoming radiation and the deduced current. For $N_{1,i}$ and $N_{2,i}$, major resonance occurs if $\omega=\pm \omega_{c_{+}}$. Also, there is another resonance at $\omega=\pm  \omega_{c_{+}}/ 2$. The deduced current in these two cases is stronger than when $\omega_{c_{+}}=0$ (Fig.~\ref{fig3} (a-b)). For $N_{3,i}$ and $N_{4,i}$, the major resonance effect is related to $\omega=\pm 2 \omega_{c_{+}}$ and then $\omega= \pm \omega_{c_{+}}$. There is also a resonance at $\omega=\pm \omega_{c_{+}}/2$. Additionally, all three resonances deduce to a current that is stronger than the case where $\omega_{c_{+}}=0$; there is no perpendicular applied magnetic field (Fig.~\ref{fig3} (c-d)).
\section{Discussion}
In discussing the strength of the effect, first, we should determine the $C$ coefficients in Eq.~\ref{deltaw}. To estimate these prefactors in phosphorene, we have substituted the values of coupling parameters, $t_i (i=1,2,...,5)$, and lattice parameters including $a_x$, $a_y$, $\alpha$, $\beta$ and $d^\prime$ in the derived change of scattering rate (Eq.~\ref{deltaw} and Fig.~\ref{fig1}(a-c)). Consequently, we can show that the $C$ coefficients are dependent on the disorder type on the top or bottom layer and on--site energies. Further details are available in Ref.~\citep{kheirabadi2020magnetic}. Accordingly, it has been shown that to have nonzero $C$ coefficients, $z\rightarrow-z$ symmetry and inter--layer symmetry should be broken; $U_1- U_2 \neq 0$. For instance, for the valence band, where the disorder is on the lower layer, $\delta= 20$ $\mathrm{m eV}$, $U_1= 0$ $\mathrm{eV}$ and $U_2=40$ $\mathrm{m eV}$, $C_1=0.034$ $\mathrm{{\AA}^{2}}$ and $C_2=-1.882 \times 10^{-8}$ $\mathrm{{\AA}^{2}}$. For the valence band, we also have $m_{xx}=0.15 m_0$, $m_{yy}=1.0 m_0$ ($m_0$ is electron free mass) \citep{lowrelax}. We also assume that $\nu_0$ is independent of energy and is equal to what has been calculated for bilayer graphene \citep{kheirabadi2016magnetic}, $B_x=B_y= 7$ $\mathrm{T}$, and inter layer distance is $d=2.13$ $\mathrm{{\AA}}$ \citep{chaves2017theoretical, pereira2015landau}. For $n_{imp}=10^{16}$ $ \mathrm{m^{-2}}$, for impurity distance $0$ $\mathrm{nm}$, according to Ref.~\citep{lowrelax}, we can assume that for all harmonics and in the $x$ and $y$ directions, the relaxation time is $0.1$ $\mathrm{ps}$. Consequently, we can show that the magnitude of the ratchet current and SHG are in the order of $\mathrm{\mu A/ cm}$; similar to that observed for monolayer graphene \citep{drexler2013magnetic}.        

Furthermore, we consider the effect of the change of inter layer asymmetry, $U_1-U_2$, on $\varphi^{\prime}$. For the valence band, when the lower layer is disordered, $\zeta=1$, $U_1=0$ $\mathrm{eV}$ and assuming $\delta=20$ $\mathrm{m eV}$, when $B_x=B_y$ and $U_2$ changes from 0 to $40$ $\mathrm{m eV}$, $\varphi^\prime$ is $-\pi/2$ with $10^{-6}$ accuracy. On the other hand, if we consider that for a constant inter layer asymmetry, $\mathbf{B_{\parallel}}$ rotates in the plane of phosphorene, where $\delta=20$ $\mathrm{m eV}$, $U_2=40$ $\mathrm{m eV}$ and the lower layer is disordered, because of the magnitude of the $C_1/C_2$, $\varphi^\prime$ obtains three values depending on the direction of the magnetic field in the plane of phosphorene. If the applied magnetic field is parallel or antiparallel to the x direction, armchair edge, the $\varphi^{\prime}$ is equal to zero. Else, it is $-\pi/2$ if $B_x B_y >0$, and is $\pi/2$ if $B_x B_y < 0$. Hence, $\varphi^\prime$ is a rectangular step function. This happens for anisotropic phosphorene because the magnitude of $C_1$ and $C_2$ that are coefficients of the magnetic field in the $y$ and $x$ directions, respectively, are different in the scattering rate; there is an in--plane anisotropy. We can also derive the same results for the conduction band \citep{kheirabadi2020magnetic}. For isotropic materials, $B_x$ and $B_y$ appear with the same magnitude of coefficients in the scattering rate \citep{kheirabadi2018cyclotron}.  

Additionally, a semi--Faraday effect has been predicted in the SHG current of isotropic materials \citep{kheirabadi2018cyclotron}. In other words, a change of direction of the in--plane magnetic field of an incoming in--plane polarized light rotates the deduced SHG current polarization direction. In anisotropic phosphorene, for linear polarized light, according to Eqs.~\ref{jxsl} and \ref{jysl}, the SHG current is dependent on $\varphi^\prime$, that is dependent on the direction of the magnetic field rather than its magnitude, and it could be $\pm \pi /2$ or $0$. Hence, it causes distinct directions for the current caused by an incoming linear polarized light.  

The magnitude of $\omega_{c_+}=e B_\perp /(2 m_+)$ for a perpendicular magnetic field, $B_\perp = 1$ $\mathrm{T}$, for electron carriers ($m_{xx}=0.15 m_0$, $m_{yy}=0.5 m_0$) is $7.1 \times 10^{11}$ $\mathrm{rad s^{-1}}$. For the cyclotron resonance condition $\omega=\omega_{c_{+}}$, this $\omega_{c_{+}}$ corresponds to a linear frequency of light $f\approx 0.11$ $\mathrm{THz}$. On the other hand, in the introduction section, we mentioned that the results of this study are valid for $\hbar \omega \leq \epsilon_f $. For $f \approx 0.11$ $\mathrm{THz}$, $\hbar \omega$ is  $0.4$ $\mathrm{m eV}$, approximately. For phosphorene, $\epsilon_f$ is $\hbar^2 \pi n/ m_d $ where n is the electron or hole density and $m_d$ is equal to $\sqrt{m_{xx}m_{yy}}$ \citep{lowrelax}. So, for carrier density $10^{16}$ $\mathrm{m^{-2}}$ \citep{kheirabadi2020magnetic}, Fermi energy $7$ $\mathrm{m eV}$ is around 20 times larger than the THz radiation energy, where the resonance current occurs. Hence, the considered semiclassical regime is correctly used in this paper.    
\section{Conclusion}
An anisotropic material like phosphorene under THz radiation deduces to three currents: first-order AC, second-order ratchet, and SHG currents. Depending on the magnitude of the perpendicular magnetic field, resonance occurs for specific laser radiation frequencies. Additionally, for an anisotropic dispersion of phosphorene, $\omega_{c_{+}}= e B_{\perp}/(2 m_+)$ is a critical factor that determines the resonance frequency, where for isotropic materials with quadratic dispersion, it is $e B_{\perp}/m$, where $m$ is the electron effective mass. 
We have also shown that anisotropy causes that instead of having a semi--Faraday effect similar to isotropic materials \citep{kheirabadi2018cyclotron}, we have a distinct direction for the SHG--related current when the magnetic field rotates in the anisotropic phosphorene plane.

\bibliographystyle{abbrv}
\bibliography{bib}

\begin{thebibliography}{10}

\bibitem{akhtar2017recent}
M.~Akhtar, G.~Anderson, R.~Zhao, A.~Alruqi, J.~E. Mroczkowska, G.~Sumanasekera,
  and J.~B. Jasinski.
\newblock Recent advances in synthesis, properties, and applications of
  phosphorene.
\newblock {\em NPJ 2D MATER APPL}, 1(1):1--13, 2017.

\bibitem{budkin2016ratchet}
G.~Budkin and S.~Tarasenko.
\newblock Ratchet transport of a two-dimensional electron gas at cyclotron
  resonance.
\newblock {\em Phys. Rev. B}, 93(7):075306, 2016.

\bibitem{butler2013progress}
S.~Z. Butler, S.~M. Hollen, L.~Cao, Y.~Cui, J.~A. Gupta, H.~R. Guti{\'e}rrez,
  T.~F. Heinz, S.~S. Hong, J.~Huang, A.~F. Ismach, et~al.
\newblock Progress, challenges, and opportunities in two-dimensional materials
  beyond graphene.
\newblock {\em ACS Nano}, 7(4):2898--2926, 2013.

\bibitem{chaves2017theoretical}
A.~Chaves, W.~Ji, J.~Maassen, T.~Dumitrica, and T.~Low.
\newblock Theoretical overview of black phosphorus.
\newblock In {\em 2D Materials: Properties and Devices}, pages 381--412.
  Cambridge University Press, 2017.

\bibitem{dantscher2015cyclotron}
K.-M. Dantscher, D.~Kozlov, P.~Olbrich, C.~Zoth, P.~Faltermeier, M.~Lindner,
  G.~Budkin, S.~Tarasenko, V.~Bel'kov, Z.~Kvon, et~al.
\newblock Cyclotron-resonance-assisted photocurrents in surface states of a
  three-dimensional topological insulator based on a strained high-mobility
  hgte film.
\newblock {\em Phys. Rev. B}, 92(16):165314, 2015.

\bibitem{das2014tunable}
S.~Das, W.~Zhang, M.~Demarteau, A.~Hoffmann, M.~Dubey, and A.~Roelofs.
\newblock Tunable transport gap in phosphorene.
\newblock {\em Nano Lett.}, 14(10):5733--5739, 2014.

\bibitem{dhanabalan2017emerging}
S.~C. Dhanabalan, J.~S. Ponraj, Z.~Guo, S.~Li, Q.~Bao, and H.~Zhang.
\newblock Emerging trends in phosphorene fabrication towards next generation
  devices.
\newblock {\em Adv. Sci.}, 4(6):1600305, 2017.

\bibitem{drexler2013magnetic}
C.~Drexler, S.~Tarasenko, P.~Olbrich, J.~Karch, M.~Hirmer, F.~M{\"u}ller,
  M.~Gmitra, J.~Fabian, R.~Yakimova, S.~Lara-Avila, et~al.
\newblock Magnetic quantum ratchet effect in graphene.
\newblock {\em Nat. Nanotechnol.}, 8(2):104--107, 2013.

\bibitem{gan2013chip}
X.~Gan, R.-J. Shiue, Y.~Gao, I.~Meric, T.~F. Heinz, K.~Shepard, J.~Hone,
  S.~Assefa, and D.~Englund.
\newblock Chip-integrated ultrafast graphene photodetector with high
  responsivity.
\newblock {\em Nat. Photonics}, 7(11):883--887, 2013.

\bibitem{geim2010rise}
A.~K. Geim and K.~S. Novoselov.
\newblock The rise of graphene.
\newblock In {\em Nanoscience and technology: a collection of reviews from
  nature journals}, pages 11--19. World Scientific, 2010.

\bibitem{glazov2014high}
M.~Glazov and S.~Ganichev.
\newblock High frequency electric field induced nonlinear effects in graphene.
\newblock {\em Phys. Rep.}, 535(3):101--138, 2014.

\bibitem{guo2016black}
Q.~Guo, A.~Pospischil, M.~Bhuiyan, H.~Jiang, H.~Tian, D.~Farmer, B.~Deng,
  C.~Li, S.-J. Han, H.~Wang, et~al.
\newblock Black phosphorus mid-infrared photodetectors with high gain.
\newblock {\em Nano Lett.}, 16(7):4648--4655, 2016.

\bibitem{huo20152d}
C.~Huo, Z.~Yan, X.~Song, and H.~Zeng.
\newblock 2d materials via liquid exfoliation: a review on fabrication and
  applications.
\newblock {\em Sci. Bull.}, 60(23):1994--2008, 2015.

\bibitem{kheirabadi2020magnetic}
N.~Kheirabadi.
\newblock Magnetic ratchet effect in phosphorene.
\newblock {\em Phys. Rev. B}, 103:045406, Jan 2021.

\bibitem{kheirabadi2016magnetic}
N.~Kheirabadi, E.~McCann, and V.~I. Fal'ko.
\newblock Magnetic ratchet effect in bilayer graphene.
\newblock {\em Phys. Rev. B}, 94(16):165404, 2016.

\bibitem{kheirabadi2018cyclotron}
N.~Kheirabadi, E.~McCann, and V.~I. Fal'ko.
\newblock Cyclotron resonance of the magnetic ratchet effect and second
  harmonic generation in bilayer graphene.
\newblock {\em Phys. Rev. B}, 97(7):075415, 2018.

\bibitem{koppens2014photodetectors}
F.~Koppens, T.~Mueller, P.~Avouris, A.~Ferrari, M.~Vitiello, and M.~Polini.
\newblock Photodetectors based on graphene, other two-dimensional materials and
  hybrid systems.
\newblock {\em Nat. Nanotechnol.}, 9(10):780--793, 2014.

\bibitem{lowrelax}
Y.~Liu, T.~Low, and P.~P. Ruden.
\newblock Mobility anisotropy in monolayer black phosphorus due to scattering
  by charged impurities.
\newblock {\em Phys. Rev. B}, 93(16):165402, 2016.

\bibitem{relaxtime}
Y.~Liu and P.~P. Ruden.
\newblock Temperature-dependent anisotropic charge-carrier mobility limited by
  ionized impurity scattering in thin-layer black phosphorus.
\newblock {\em Phys. Rev. B}, 95(16):165446, 2017.

\bibitem{luo2015microfiber}
Z.-C. Luo, M.~Liu, Z.-N. Guo, X.-F. Jiang, A.-P. Luo, C.-J. Zhao, X.-F. Yu,
  W.-C. Xu, and H.~Zhang.
\newblock Microfiber-based few-layer black phosphorus saturable absorber for
  ultra-fast fiber laser.
\newblock {\em Opt. Express}, 23(15):20030--20039, 2015.

\bibitem{novoselov2012roadmap}
K.~S. Novoselov, V.~Fal'ko, L.~Colombo, P.~Gellert, M.~Schwab, K.~Kim, et~al.
\newblock A roadmap for graphene.
\newblock {\em Nature}, 490(7419):192--200, 2012.

\bibitem{novoselov2004electric}
K.~S. Novoselov, A.~K. Geim, S.~V. Morozov, D.~Jiang, Y.~Zhang, S.~V. Dubonos,
  I.~V. Grigorieva, and A.~A. Firsov.
\newblock Electric field effect in atomically thin carbon films.
\newblock {\em Science}, 306(5696):666--669, 2004.

\bibitem{olbrich2016terahertz}
P.~Olbrich, J.~Kamann, M.~K{\"o}nig, J.~Munzert, L.~Tutsch, J.~Eroms, D.~Weiss,
  M.-H. Liu, L.~Golub, E.~Ivchenko, et~al.
\newblock Terahertz ratchet effects in graphene with a lateral superlattice.
\newblock {\em Phys. Rev. B}, 93(7):075422, 2016.

\bibitem{olbrich2013giant}
P.~Olbrich, C.~Zoth, P.~Vierling, K.-M. Dantscher, G.~Budkin, S.~Tarasenko,
  V.~Bel'kov, D.~Kozlov, Z.~Kvon, N.~Mikhailov, et~al.
\newblock Giant photocurrents in a dirac fermion system at cyclotron resonance.
\newblock {\em Phys. Rev. B}, 87(23):235439, 2013.

\bibitem{pereira2015landau}
J.~Pereira~Jr and M.~Katsnelson.
\newblock Landau levels of single-layer and bilayer phosphorene.
\newblock {\em Phys. Rev. B}, 92(7):075437, 2015.

\bibitem{peruzzini2019perspective}
M.~Peruzzini, R.~Bini, M.~Bolognesi, M.~Caporali, M.~Ceppatelli, F.~Cicogna,
  S.~Coiai, S.~Heun, A.~Ienco, I.~I. Benito, et~al.
\newblock A perspective on recent advances in phosphorene functionalization and
  its applications in devices.
\newblock {\em Eur. J. Inorg. Chem.}, 2019(11-12):1476--1494, 2019.

\bibitem{shim2017electronic}
J.~Shim, H.-Y. Park, D.-H. Kang, J.-O. Kim, S.-H. Jo, Y.~Park, and J.-H. Park.
\newblock Electronic and optoelectronic devices based on two-dimensional
  materials: from fabrication to application.
\newblock {\em Adv. Electron. Mater.}, 3(4):1600364, 2017.

\bibitem{sun2016optical}
Z.~Sun, A.~Martinez, and F.~Wang.
\newblock Optical modulators with 2d layered materials.
\newblock {\em Nat. Photonics}, 10(4):227--238, 2016.

\bibitem{viti2015black}
L.~Viti, J.~Hu, D.~Coquillat, W.~Knap, A.~Tredicucci, A.~Politano, and M.~S.
  Vitiello.
\newblock Black phosphorus terahertz photodetectors.
\newblock {\em Adv. Mater.}, 27(37):5567--5572, 2015.

\bibitem{xia2014rediscovering}
F.~Xia, H.~Wang, and Y.~Jia.
\newblock Rediscovering black phosphorus as an anisotropic layered material for
  optoelectronics and electronics.
\newblock {\em Nat. Commun.}, 5(1):1--6, 2014.

\bibitem{AsgariHamil}
M.~Zare, B.~Z. Rameshti, F.~G. Ghamsari, and R.~Asgari.
\newblock Thermoelectric transport in monolayer phosphorene.
\newblock {\em Phys. Rev. B}, 95(4):045422, 2017.

\end{thebibliography}

\end{document}